\documentclass[aps,pra,showpacs,twocolumn,superscriptaddress]{revtex4-1}
\usepackage{amsfonts,amsmath,amssymb,graphicx}
\usepackage[usenames,dvipsnames]{xcolor}
\usepackage{pdfpages}
\usepackage{footmisc}
\usepackage[normalem]{ulem}
\usepackage{color}

\begin{document}

\title{Decoherence of a single spin coupled to an interacting spin bath}

\author{Ning Wu}
\email{wun1985@gmail.com; \emph{Present address}: IFIMAC, Universidad Aut\'onoma de Madrid, E-28049 Madrid, Spain}

\affiliation{Department of Chemistry, Princeton University, Princeton, New Jersey 08544, USA}
\author{Nina Fr\"ohling}
\affiliation{Lehrstuhl f\"ur Theoretische Physik II, Technische Universit\"at Dortmund, 44221 Dortmund, Germany}
\author{Xi Xing}
\affiliation{Department of Chemistry, Princeton University, Princeton, New Jersey 08544, USA}

\author{Johannes Hackmann}
\affiliation{Lehrstuhl f\"ur Theoretische Physik II, Technische Universit\"at Dortmund, 44221 Dortmund, Germany}
\author{Arun Nanduri}
\affiliation{Department of Chemistry, Princeton University, Princeton, New Jersey 08544, USA}
\author{Frithjof B. Anders}
\email{frithjof.anders@tu-dortmund.de}
 \affiliation{Lehrstuhl f\"ur Theoretische Physik II, Technische Universit\"at Dortmund, 44221 Dortmund, Germany}
\author{Herschel Rabitz}
\email{hrabitz@princeton.edu}
\affiliation{Department of Chemistry, Princeton University, Princeton, New Jersey 08544, USA}


\begin{abstract}
Decoherence of a central spin coupled to an interacting spin bath via
inhomogeneous Heisenberg coupling is studied by two different
approaches, namely an exact equations of motion (EOMs) method and a
Chebyshev expansion technique (CET).  By assuming a wheel topology of
the bath spins with uniform nearest-neighbor $XX$-type intrabath
coupling, we examine the central spin dynamics with the bath prepared
in two different types of bath initial conditions. For fully
polarized baths in strong magnetic fields, the polarization dynamics
of the central spin exhibits a collapse-revival behavior in the
intermediate-time regime. Under an antiferromagnetic bath initial
condition, the two methods give excellently consistent central spin
decoherence dynamics for finite-size baths of $N\leq14$ bath spins.
The decoherence factor is found to drop off abruptly on a short time
scale and approach a finite plateau value which depends on the
intrabath coupling strength non-monotonically. In the ultrastrong
intrabath coupling regime, the plateau values show an oscillatory
behavior depending on whether $N/2$ is even or odd. The observed
results are interpreted qualitatively within the framework of the EOM
and perturbation analysis. The effects of anisotropic spin-bath coupling and inhomogeneous intrabath bath couplings are briefly discussed. Possible experimental realization of the model in a modified quantum corral setup is suggested. 

\end{abstract}


\maketitle

\section{Introduction}

A single central spin coupled to a spin environment~\cite{Stamp} is a widely studied quantum many-body system because of its high relevance to solid-state based quantum computation~\cite{pola1,pola2,EOM,master1,master2,master3,4th,effective1,effective2,effective3,semi1,semi1.5,semi2,BA1,pola3,BA2,BA3,
Exact1,Exact2,Exact4,mc1,mc2,anders,Uhrig,FischerLoss2008,Testelin2009,Merkulov2002,natmat,XY,UhrigHackmann2014,isingsb}, quantum decoherence and quantum information~\cite{intra1,intra3,intra4,PRA,bose,lukin,lai,discord,2008science,nat}, and excitation energy transfer~\cite{johnson,prl2012,JCP}, etc. In these systems, a single spin is inevitably coupled to its surrounding (pseudo-)spin environment and it is important to understand and control the decoherence of the central spin induced by environmental fluctuations. Among these, the central spin or Gaudin model~\cite{Gaudin} plays an important role in describing the decoherence dynamics of an electron trapped in a quantum dot, which has long been proposed to be a promising candidate for realizing a qubit for solid state quantum computation~\cite{loss1998,rmp2006}.  The trapped central spin $\mathbf{S}$ coupled to a bath of $N$ nuclear spins $\{\mathbf{S}_j\}$ via inhomogeneous hyperfine couplings $\{g_j\}$ is well modeled by the Hamiltonian
\begin{eqnarray}\label{CSM}
H=\omega S_z+2\sum^N_{j=1}g_j \mathbf{S}\cdot \mathbf{S}_j,
\end{eqnarray}
where $\omega=g\mu_B B$ is the Larmor frequency of the central spin due to the externally applied magnetic field $B$.

\par The dynamics of the central spin model Eq. (\ref{CSM}) has been studied theoretically with a wide variety of methods. In the strong magnetic field limit $\omega>N\max\{g_j\}$, perturbative treatments~\cite{pola1,pola2,EOM}, non-Markovian master equation methods~\cite{master1,master2,master3,4th}, and effective Hamiltonian approaches~\cite{effective1, effective2,effective3} in terms of the flip-flop hyperfine interaction $\sum_jg_j(S_+S^-_j+S_-S^+_j)$ were commonly employed. In the low-field limit, semiclassical approaches based on the spin coherent state representation~\cite{semi1,semi1.5,semi2}, exact methods via either the Bethe ansatz techniques~\cite{BA1,pola3,BA2,BA3} or exact diagonalization~\cite{Exact1,Exact2,Exact4} have been used intensively to study the dynamics of the model.
Most recent advances include a combination of algebraic Bethe ansatz and Monte Carlo simulation~\cite{mc1,mc2}, as well as an advanced numerical method based on the Chebyshev polynomial technique~\cite{anders}.
The extension of the density matrix renormalization group \cite{Schollwoeck2011}
to the real-time dynamics in central spin model~\cite{Uhrig} can treat up to $N\approx 1000$ nuclear spin,
but is restricted to rather short times.
\par In the above studies, the dipolar interaction between nuclear spins was often neglected because it starts to play a role only on longer time scales compared to the hyperfine interaction. However, these effects need to be taken into account if one is interested in the long-time behavior of the central spin dynamics. On the other hand, it is also of interest to study the influence of intrabath interactions~\cite{UhrigHackmann2014,isingsb,intra1,intra3,intra4,PRA} on the decoherence properties of the central spin from a purely theoretical point of view. The bath spins described by Eq.~(\ref{CSM}) form a noninteracting spin cluster whose spatial correlations are only reflected from its coupling $\{g_j\}$ to the central spin.
\par In this work, we use exact equations of motion approach combined with a Chebyshev expansion technique to study the decoherence dynamics of a central spin coupled to a periodic $XX$ spin chain with nearest-neighbor interaction. Such a system can be obtained by arranging the bath spins uniformly on a ring but leaving the hyperfine interactions unchanged, then by turning on the homogeneous
$XX$-type interaction between adjacent bath spins. We note that such a fictitious one dimensional structure of the spin bath also helped implement density matrix renormalization group algorithms of Eq.~(\ref{CSM})~\cite{Uhrig}. The present system can also be thought of as an extension to the model studied in our previous work~\cite{PRA}, where we obtained the exact reduced dynamics of a central spin coupled to an
$XX$ spin ring via the $XX$-type interactions as well. Taking advantage of the translational invariance of the spin bath as well as the conservation of the total magnetization, we can write down the evolved state of the system in terms of Jordan-Wigner fermions in the momentum space of the chain. Interesting quantities like the decoherence factor and polarization of the central spin then can be expressed as combinations of the corresponding time-dependent probability amplitudes.
\par We have studied the spin decay of an initially fully polarized  central spin as well as the
decoherence factor \cite{zurek}. For the central spin model described by Eq.~(\ref{CSM}), due to the integrability and the conservation of total magnetization in $z$-direction, a partial coherence is maintained \cite{mc1,mc2,Uhrig,anders,UhrigHackmann2014}
depending on the initial
preparations of the system \cite{mc1,mc2}. Switching on the $XX$-ring coupling decreases
the long-time limit of the decoherence factor and, therefore, the partial coherence in the system.
Once the $XX$-ring coupling exceeds some value, however, we find a gradual increase of coherence, which appears to be counterintuitive at first. We present a physical picture for explaining these surprising findings.
\par Although the EOM method employed here restrict us to small baths with $N\leq 14$, it provides benchmarks for testing the CET method,
which can be applied to large baths. It has been demonstrated~\cite{anders} that the CET approach \cite{TalEzer-Kosloff-84,Kosloff-94,Exact3,ZangHarmon2006,Fehske-RMP2006} is well suited to access the quantitive feature of the central spin dynamics. In particular, quantum corrections \cite{master1,FischerLoss2008} to a quasi-classical treatment of the nuclear spin bath \cite{Merkulov2002} have been revealed. Furthermore, the EOM approach also provides a possibly intuitive way to understand the observed nontrivial dependence of the decoherence on the intrabath coupling strengths. We finally discuss potential reality of the design in the laboratory.
\par The rest of the paper is structured as follows. In Sec. II we introduce our model Hamiltonian and describe the EOM approach and the CET technique in detail. In Sec. III we study the central spin polarization for a fully polarized bath, and the decoherence dynamics for a zero-polarization bath initial state by employing the two methods. Conclusions are drawn in Sec. IV.
\section{Methodology}
\subsection{The interacting central spin model}

By arranging the $N$ spins ($N$ even) in the spin bath into a ring-like fashion, we consider an extension of model (\ref{CSM}) by introducing a homogeneous nearest-neighbor $XX$-type interaction between adjacent bath spins. Such a system can be described by the Hamiltonian
\begin{eqnarray}\label{H}
H&=&H_{\rm{S}}+H_{\rm{B}}+H_{\rm{SB}},\nonumber\\
H_{\rm{S}}&=&\omega S_z,\nonumber\\
H_{\rm{B}}&=&J\sum^N_{j=1}(S^x_jS^x_{j+1}+S^y_jS^y_{j+1}),\nonumber\\
H_{\rm{SB}}&=&2\sum^N_{j=1}[ g_j(S_xS^x_j+S_yS^y_j)+g'_jS_zS^z_j],
\end{eqnarray}
where we have separated $H$ into the system part (central spin) $H_{\rm{S}}$, the environment part ($XX$ ring) $H_{\rm{B}}$, and the anisotropic Heisenberg system-bath coupling $H_{\rm{SB}}$ between the two.
In Eq.~(\ref{H}), $S_\alpha$ and $S^\alpha_j$ are the spin-1/2 operators for the central spin and spin $j$ in the $XX$
bath, respectively, $\omega$ is the Larmor frequency in an external magnetic field,
$J$ is the uniform nearest-neighbor intrabath coupling strength, and $\{g_j\}$ as well as $\{g'_j\}$ are the exchange interactions between the central spin and the bath spins.

The Hamiltonian (\ref{H}) includes the anisotropic spin model \cite{FischerLoss2008,Testelin2009}
which corresponds to $\Lambda = g'_j/g_j$ and is relevant to quantum dots charged with a single hole, where $\Lambda$ depends on the mixing
of heavy and light holes. For $\Lambda=1$, the isotropic model describing electron charged quantum dots is recovered. The Hamiltonian (\ref{H}) also extends the model proposed in Ref.~\cite{PRA} by including the Ising component of the system-bath coupling, which is essential for the model to cover the Gaudin model in the limit of $J\to0$. It is thus experimentally more relevant to introduce the Ising part of the system-bath coupling. As we will see, the decay of the central spin decoherence can be suppressed by increasing $\Lambda$.

\par It can be easily seen that the total magnetization $M=S_z+L_z$ is conserved, where $L_z=\sum_iS^z_i$ is the magnetization of the spin bath. Thus, we dropped the Zeeman term for the bath spins in Eq. (2). However, the model described Eq. (\ref{H}) is only integrable for $J=0$ and $\Lambda=1$,
i.e., for the conventional Gaudin model with a noninteracting spin bath and isotropic spin-bath coupling. Hence, results from the Bethe ansatz solution generally cease to be applicable here.

\par In order to separate dynamics governed by the fluctuations of the Overhauser field \cite{Merkulov2002}
from the dynamics induced by the spin-bath interaction, we define
the energy scale $\omega_{\rm{fluc}}$~\cite{anders}
 \begin{eqnarray}
 \label{eq:w-fluc}
\omega_{\rm{fluc}}=2\sqrt{\sum^N_{j=1}g^2_j}
 \end{eqnarray}
associated by the fluctuation of the Overhauser field. The such measured short-time dynamics will be universal for different number of bath spins, which can be physically understood from second-order short-time perturbation of the dynamics. Regardless of the initial conditions, the second-order process flipping the same spin back always yields a dimensionless term $\propto  (\sum^N_{j=1}g^2_j) t^2$. Below, we will measure all energies in units of  $\omega_{\rm{fluc}}$.

\subsection{The exact equations of motion approach}

The one-dimensional bath Hamiltonian $H_{\rm B}$ can be diagonalized by using the Jordan-Wigner transformation $S^-_i=S^x_i-iS^y_i=\prod^{i-1}_{j=1}(1-2c^\dag_jc_j)c_i,~S^z_i=c^\dag_ic_i-\frac{1}{2}$ with $c^\dag_i$ a fermion creation operator. Then $H$ describes a central spin immersed in a spinless fermion bath
\begin{eqnarray}\label{H_jw}
H&=&\omega' S_z+\frac{J}{2}\sum^N_{j=1}(c^\dag_jc_{j+1}+c^\dag_{j+1}c_j) \nonumber\\
&&+ \sum^N_{j=1}[g_j(c^\dag_jT_jS_-+c_jT^\dag_jS_+)+2g'_jS_z c^\dag_jc_j ],
\end{eqnarray}
where the Jordan-Wigner string $T_j=\exp(i\pi\sum^{j-1}_{l=1}c^\dag_lc_l)$ and
\begin{eqnarray}\label{wp}
\omega'=\omega-\sum^N_{j=1} g'_j
\end{eqnarray}
is the effective magnetic field experienced by the central spin when all the bath spins are pointing down. Depending on the parity of the eigenvalue $N_f$ of the total fermion number operator $\mathcal{N}_f=\sum^{N}_{l=1}c^\dag_lc_l$, two subpaces with projection operators $P_+=\frac{1+T_{N-1}}{2}$ (for even $N_f$) and $P_-=\frac{1-T_{N-1}}{2}$ (for odd $N_f$) can be identified. This results in antiperiodic $c_{N+1}=-c_1$ or periodic boundary conditions $c_{N+1}=c_1$ imposed on the fermions for even or odd $N_f$. As a result, two sets of Fourier transformations can be introduced
\begin{eqnarray}
\label{eq:6}
c_j=\frac{1}{\sqrt{N}}\sum_{k\in K_\pm}e^{ikj}c_{\pm,k},
\end{eqnarray}
where $\{c_{\pm,k}\}$ are Fourier modes with wave numbers surviving in $K_\pm=\{-\pi+(4j-3)\frac{\pi}{2N}\pm\frac{\pi}{2N}\}$, $j=1,...,N$. Now $H_{\rm B}$ is diagonalized as
\begin{eqnarray}
H_{\rm B}&=&\sum_{\sigma=\pm}P_\sigma H_\sigma P_\sigma,\nonumber\\
H_\sigma&=&\sum_{k\in K_\sigma}\varepsilon_k c^\dag_{\sigma,k}c_{\sigma,k},
\end{eqnarray}
where $\varepsilon_k=J\cos k$ is the single-particle spectrum. We will work in the interaction picture with respect to the first two terms of Eq. (\ref{H_jw}). Direct calculation gives
\begin{eqnarray}
&&H_I(t)=\sum^N_{j=1}g_j(S_- e^{-i\omega' t} \sum_{\sigma=\pm}P_\sigma e^{iH_\sigma t}c^\dag_j T_j e^{-iH_{-\sigma}t}P_{-\sigma} \nonumber\\
&&+\textrm{h.c.})+2\sum^N_{j=1}g'_jS_z\sum_{\sigma=\pm} P_\sigma e^{iH_\sigma t}c^\dag_jc_j e^{-iH_\sigma t}P_\sigma.
\end{eqnarray}
We assume that initially the central spin and the bath are decoupled (the up and down states of each spin are denoted by $|1\rangle$ and $|\bar{1}\rangle$, respectively)
\begin{eqnarray}
\label{eq:psi-initial}
|\psi(0)\rangle&=&(a_{\bar{1}}|\bar{1}\rangle+a_1|1\rangle)|\phi_b\rangle,
\end{eqnarray}
where the coefficients satisfy $|a_{\bar{1}}|^2+|a_{1}|^2=1$ and $|\phi_b\rangle$ is the bath initial state which is assumed to have fixed magnetization $L_z=m-\frac{N}{2}$ with $m$ the number of up spins in the bath. For $|\phi_b\rangle$ having nonzero overlap in several $L_z$-subspaces the time evolution obtained in distinct subspaces can be superimposed.

It is convenient to define two vectors made up of momentum and spatial indices
\begin{eqnarray}
\vec{k}_m\equiv(k_1,k_2,...,k_m),~\vec{j}_m\equiv(j_1,j_2,...,j_m),
\end{eqnarray}
with the convention $k_1<k_2<...<k_m$ and $1\leq j_1<j_2<...<j_m\leq N$, and the total energy of the $m$ fermions occupying the set of modes $\vec{k}_m$
\begin{eqnarray}
\label{eqn:def-Ecal}
\mathcal{E}_{\vec{k}_m}\equiv\sum^m_{l=1}\varepsilon_{k_l}.
\end{eqnarray}
Then $|\phi_b\rangle$ can be written in the momentum space
\begin{eqnarray}\label{ini}
|\phi_b\rangle=\sum_{ \vec{k}_m}\chi_{\vec{k}_m}|\vec{k}_{\sigma,m}\rangle,
\end{eqnarray}
where $|\vec{k}_{\sigma,m}\rangle\equiv \prod^m_{l=1}c^\dag_{\sigma,k_l}|0\rangle$ with $\sigma=+(-)$ for even (odd) $m$. Here $|0\rangle$ is the vacuum state of the fermions corresponding to the fully polarized bath state $|\bar{1}...\bar{1}\rangle$. Normalization of $|\phi_b\rangle$ gives
\begin{eqnarray}
 \sum_{ \vec{k}_m}|\chi_{\vec{k}_m}|^2=1.
\end{eqnarray}

Since the total magnetization is conserved, the time evolved state in the interaction picture is of the form
\begin{eqnarray}\label{psiI}
&&|\psi_I(t)\rangle=\nonumber\\
&&~~\sum_{\vec{k}_m}[a_{\bar{1}}A_{\vec{k}_m}(t)|\bar{1}\rangle+a_1B_{\vec{k}_m}(t)|1\rangle]e^{i\mathcal{E}_{\vec{k}_m}t}|\vec{k}_{\sigma,m}\rangle\nonumber\\
&&~~+\sum_{\vec{k}_{m+1}}a_1D_{\vec{k}_{m+1}}(t)|\bar{1}\rangle e^{i\mathcal{E}_{\vec{k}_{m+1}}t}|\vec{k}_{-\sigma,m+1}\rangle\nonumber\\
&&~~+\sum_{\vec{k}_{m-1}}a_{\bar{1}}C_{\vec{k}_{m-1}}(t)|1\rangle e^{i\mathcal{E}_{\vec{k}_{m-1}}t}|\vec{k}_{-\sigma,m-1}\rangle.
\end{eqnarray}
The nonvanishing initial values of $A$, $B$, $C$, and $D$ can be read from Eq. (\ref{ini}) as
\begin{eqnarray}
A_{\vec{k}_m}(0)=B_{\vec{k}_m}(0)=\chi_{\vec{k}_m}.
\end{eqnarray}
Applying the Schr\"odinger equation $i\partial_t|\psi_I(t)\rangle=H_I(t)|\psi_I(t)\rangle$ and after some algebra we arrive at the following sets of EOMs for the probability amplitudes $A$, $B$, $C$, and $D$
\begin{eqnarray}\label{C}
i\dot{C}_{\vec{p}_{m-1}}&=&\mathcal{E}_{\vec{p}_{m-1}} C_{\vec{p}_{m-1}}+(  \sum_{ \vec{k}_{m}}A_{ \vec{k}_{m}} e^{i \omega' t} f_{ \vec{k}_{m}; \vec{p}_{m-1}}\nonumber\\
&& +\sum_{\vec{k}_{m-1}}C_{ \vec{k}_{m-1}}   f'_{ \vec{k}_{m-1}; \vec{p}_{m-1}}),
\end{eqnarray}
\begin{eqnarray}\label{A}
 i\dot{A}_{\vec{p}_{m }}&=& \mathcal{E}_{\vec{p}_{m }}  A_{\vec{p}_{m }}+(\sum_{\vec{k}_{m-1}}C_{\vec{k}_{m-1}}e^{-i \omega' t} f^*_{\vec{p}_{m}; \vec{k}_{m-1}} \nonumber\\
&& -\sum_{\vec{k}_{m}}A_{\vec{k}_{m }}  f' _{\vec{k}_{m }; \vec{p}_{m }}) ,
\end{eqnarray}
and
\begin{eqnarray}\label{B}
 i\dot{B}_{ \vec{p}_{m }}&=& \mathcal{E}_{\vec{p}_{m }}  B_{\vec{p}_{m }}+(\sum_{\vec{k}_{m+1}}D_{\vec{k}_{m+1}}e^{i \omega' t} f_{\vec{k}_{m+1}; \vec{p}_{m }} \nonumber\\
&&+\sum_{\vec{k}_{m}}B_{ \vec{k}_{m }} f'_{ \vec{k}_{m} ; \vec{p}_{m }}),
\end{eqnarray}
\begin{eqnarray}\label{D}
 i\dot{D}_{\vec{p}_{m+1}}&=&\mathcal{E}_{\vec{p}_{m+1 } }D_{\vec{p}_{m+1}}+ ( \sum_{\vec{k}_{m}}B_{\vec{k}_{m}}e^{-i \omega' t}f^*_{\vec{p}_{m+1}; \vec{k}_{m} } \nonumber\\
&&-\sum_{\vec{k}_{m+1}}D_{\vec{k}_{m+1}} f'_{\vec{k}_{m+1}; \vec{p}_{m+1}} ).
\end{eqnarray}
The derivation of these EOMs for the case of $g'_j=0$ can be found in Ref.~\cite{PRA} and extension to the present model is straightforward. The auxiliary functions $f$ and $f'$ here are defined to be
\begin{eqnarray}\label{ff}
f_{\vec{k}_{n+1}; \vec{p}_n}(\{g_j\})=\sum_{ \vec{j}_{n+1}}   S_{ \vec{k}_{n+1}; \vec{j}_{n+1}}  \sum^{n+1}_{l=1}g_{j_l}S^*_{\vec{p}_n; \vec{j}^{(l)}_{n+1}},
\end{eqnarray}
and
\begin{eqnarray}\label{ffp}
f'_{\vec{k}_{n }; \vec{p}_n}(\{g'_j\})=\sum_{ \vec{j}_{n }}   S_{\vec{k}_{n }; \vec{j}_{n }} S^*_{ \vec{p}_n; \vec{j}_n}(\sum^{n }_{l=1}g'_{j_l}).
\end{eqnarray}
where the Slater determinant $S_{\vec{k}_{n }; \vec{j}_{n }}=\det (O)$ with the $n\times n$ matrix $O$ made up of plane waves
\begin{eqnarray}\label{Slater}
O_{a,b}=\frac{1}{\sqrt{N}} e^{ik_a j_b},~(a,b=1,2,...,n)
\end{eqnarray}
The vector $\vec{j}^{(l)}_{n+1}=(j_1,...,j_{l-1},j_{l+1},...,j_{n+1})$ in Eq. (\ref{ff}) is a string of length $n$ with the element $j_l$ being removed from $\vec{j}_{n+1}$.

We now turn to some remarks regarding the above procedures. (i) We note that our treatment here is similar to that in Ref.~\cite{XY} where the real-space basis states of the bath were used to study the dynamics of model~(\ref{CSM}); (ii) The system-bath coupling configurations $\{g_j\}$ ($\{g'_j\}$) are incorporated into the $f$-functions ($f'$-functions), while the intrabath coupling $J$ enters the equations of motion through the occupation energy $\mathcal{E}_{\vec{p}_m}$. Furthermore, the time-dependent phase factors $e^{\pm i\omega't}$ associated with the $f$-functions depend both on the bare magnetic field $\omega$ and average of the transverse couplings $\{g'_j\}$; (iii) The number of $f$ or $f'$ functions is of the order of $\sim (C^m_{N})^2$ (here $C^m_N=\frac{N!}{(N-m)!}$ is the binomial coefficient), as a result, it is usually time consuming to numerically calculate these functions for $m\sim N/2$; (iv) Intriguingly, the same structure with the $f$-functions in Eq.~(\ref{ff}) also appeared in the study of nonlinear optical response of one-dimensional molecular aggregates~\cite{Spano}.

We are interested in the polarization $\langle S_z(t)\rangle$ and  the decoherence factor $|r(t)|=|\langle S_+(t)\rangle/\langle S_+(0)\rangle|$ \cite{zurek}, which can be easily calculated by transforming Eq.~(\ref{psiI}) back to the Schr\"odinger picture
\begin{eqnarray}
\langle S_z(t)\rangle&=& \frac{1}{2} |a_{\bar{1}}|^2(\sum_{\vec{k}_{m-1}}|C_{\vec{k}_{m-1}}|^2-\sum_{\vec{k}_{m }}|A_{\vec{k}_{m }}|^2)\nonumber\\
&+&\frac{1}{2} |a_{1}|^2(\sum_{\vec{k}_{m }}|B_{\vec{k}_{m}}|^2-\sum_{\vec{k}_{m+1 }}|D_{\vec{k}_{m+1 }}|^2),
\\
\label{eq:r-vs-t}
|r(t)|&=&|\sum_{\vec{k}_m} A_{\vec{k}_m}(t) B^*_{\vec{k}_m}(t)|.
\end{eqnarray}
Using Eq.~(\ref{psiI}), it can be checked that $\langle S_z(t)\rangle$ is two times the autocorrelation function~\cite{pola2,Uhrig}
 \begin{eqnarray}\label{cf}
S(t)=\langle\psi(0)|S_z(t)S_z|\psi(0)\rangle=\frac{1}{2}\langle S_z(t)\rangle.
 \end{eqnarray}
Furthermore, as a direct consequence of the EOMs Eqs.~(\ref{C})-(\ref{D}), one can derive the following relation
 \begin{eqnarray}\label{gaussian}
\frac{d|r(t)|}{dt}|_{t=0}=0,
 \end{eqnarray}
 regardless of the bath initial conditions. Eq.~(\ref{gaussian}) states that the short-time decay of $|r(t)|$ begins at least with the quadratic term of $t$, e.g., a Gaussian initial decay of $|r(t)|$ was observed in Ref.~\cite{PRA}.
\par Before ending this section, let us see how the EOMs look like for a uniform hyperfine coupling. In this special case, we have $g_j=g$ and $g'_j=g'$, then the $f$- and $f'$- functions Eqs.~(\ref{ff}) and (\ref{ffp}) can be evaluated in closed form~\cite{Hanamura}
\begin{eqnarray}\label{ff_cf}
f_{\vec{k}_{n+1}; \vec{p}_n}&&(g)=2^ngN^{\frac{1}{2}-n} \delta(\sum^{n+1}_{j=1}k_j,\sum^n_{i=1}p_i)\\
&&
\times
\frac{\prod_{i>i'}(e^{-ip_i}-e^{-ip_{i'}})\prod_{j>j'}(e^{ik_j}-e^{ik_{j'}})}{\prod^n_{i=1}\prod^{n+1}_{j=1}(1-e^{-i(k_j-p_i)})},
\nonumber
\end{eqnarray}
and
\begin{eqnarray}\label{ffp_cf}
f'_{\vec{k}_{n }; \vec{p}_n}(g')=ng' \prod^n_{i=1}\delta_{k_i,p_i}.
\end{eqnarray}
Eqs.~(\ref{ff_cf}) and (\ref{ffp_cf}) show that the spin-flip part of the anisotropic hyperfine coupling will drive the initial state $|\psi_I(0)\rangle=\sum_{\vec{k}_m}(a_{\bar{1}}|\bar{1}\rangle+a_1|1\rangle)\chi_{\vec{k}_m}|\vec{k}_{\sigma,m}\rangle$ into states within the $(m\pm 1)$-sectors that conserve the total initial momentum $\sum^m_{i=1}k_{i}$, while the Ising component only induces transitions into the states within the same $m$-sector that preserve the exact momentum configuration $\vec{k}_m$. This arises from the circular symmetry of the interaction configuration for the uniform hyperfine coupling. For general nonuniform coupling, however, no analytical results for the $f$- and $f'$-functions exist and we have to resort to numerical methods.
\subsection{The Chebyshev expansion technique}
The Chebyshev expansion technique (CET)
\cite{TalEzer-Kosloff-84,Kosloff-94,Exact3,Fehske-RMP2006} has been
developed for more than three decades and offers an accurate way to calculate the
time evolution of a general single initial state $|\psi_0 \rangle$ under the
influence of a general time-independent Hamiltonian $H$
with a bound spectrum:
\begin{eqnarray}
|\psi(t) \rangle &=& e^{-i H t}|\psi_0\rangle .
\end{eqnarray}
The main idea of the method is to construct a stable
numerical approximation for the time-evolution operator $e^{-i H t}$
that is independent of the initial state $|\psi_0\rangle$,
whose error can be reduced to machine precision for any given time
$t$. While this can be achieved by any complete set of orthogonal
functions, the  Chebyshev polynomials guarantee a linear dependence
of the time $t$ and the order of the expansion.

If the spectrum of the Hamiltonian is bound to $E_{\rm min}\le E\le
E_{\rm max}$, the time-evolution operator $e^{-i  H t}$ can be
expanded  as
\begin{equation}
e^{-i H t} = \sum_{n = 0}^{\infty} b_n(t) T_n(H') \,
\label{Chebyshev-exp-e^-iH}
\end{equation}
after mapping the Hamiltonian to the
dimensionless $H' = (H-\alpha)/\Delta E$ where we have defined the
center of the energy spectrum $\alpha = (E_{\rm max}+E_{\rm
min})/2$ and its half-width $\Delta E = (E_{\rm max}-E_{\rm
min})/2$. The time dependency is entering only the analytically
known expansion coefficients
\begin{eqnarray}
\label{eqn:b_n} b_n(t)&=& (2-\delta_{0,n} ) i^n e^{-i\alpha t} J_n (
\Delta E t),
\end{eqnarray}
where $J_n(x)$ is the Bessel function of $n$-th order, while the dynamics of the systems is included in the Chebyshev polynomials
$T_n(z)$ defined by the recursion relation
\begin{equation}
T_{n+1}(z) = 2zT_n(z)-T_{n-1}(z) ,
\end{equation}
 subject to the initial conditions $T_0(z) = 1$ and
$T_1(z) = z$.
By applying the expansion of time-evolution operator
\eqref{Chebyshev-exp-e^-iH} to the initial state
$|\psi_0\rangle$ one obtains
\begin{equation}
|\psi(t) \rangle = \sum_{n = 0}^{\infty} b_n(t)
|\phi_n \rangle ,
\label{CET}
\end{equation} where the infinite set of states $|\phi_n \rangle =
T_n(H') |\psi_0\rangle$ obey the recursion
relation\cite{Fehske-RMP2006}
\begin{equation}
|\phi_{n+1} \rangle = 2H' |\phi_n \rangle -
|\phi_{n-1} \rangle ,
\label{chebyshev-recursion-relation}
\end{equation}
subject to the initial condition $|\phi_0 \rangle =
|\psi_0 \rangle$ and $|\phi_1 \rangle = H' |\psi_0 \rangle$.
\par The Chebyshev recursion relation of
Eq.~(\ref{chebyshev-recursion-relation}) reveals the iterative nature
of the calculations. Starting from the initial state $|\psi_0\rangle$,
one constructs all subsequent states $|\phi_n\rangle$ using repeated
applications of the dimensionless Hamiltonian $H'$.
Since $J_n(x) \sim (e x/2 n)^n$ for large order $n$, the Chebyshev
expansion converges quickly as $n$ exceeds $\Delta E t$. This allows
to terminate the series (\ref{CET}) after a finite number of elements
$N_C$ guaranteeing an exact result up to a well defined order.
More details on the method can be found in Refs. \cite{anders,Exact3,ZangHarmon2006}
and in the review on kernel polynomial methods \cite{Fehske-RMP2006}.

\section{Results}
\subsection{Fully polarized bath: application of the EOM approach}
As an illustration for the EOM approach, we first apply it to the simplest initial bath configuration, the fully polarized bath~\cite{pola1,pola2,pola3},
 \begin{eqnarray}\label{fullp}
|\phi_b\rangle&=& |1...1\rangle
=\prod^{N}_{l=1}c^\dag_l |0\rangle= S^*_{\vec{k}_{N};\vec{j}_N} |\vec{k}_{+,N}\rangle.
 \end{eqnarray}
The central spin is assumed to be in its down state $|\bar{1}\rangle$ initially, i.e., $a_{\bar{1}}=1$ and $a_1=0$, so that only $A$ and the $N$ $C$'s are used.
Eq.~(\ref{fullp}) yields the following initial conditions for the coefficients
\begin{eqnarray}\label{AC_ini}
A_{\vec{p}_{N}}=S^*_{\vec{p}_{N};\vec{j}_N},~C_{\vec{p}_{N-1}}=0.
 \end{eqnarray}
For the fully polarized bath, the dimension of the Hilbert space involved is $N+1$ which allows for relatively large number of bath spins. To compare our results with prior works, we follow Ref.~\cite{pola3} to choose the following distribution of $\{g_j\}$
\begin{eqnarray}
g_j=\frac{g}{N}e^{-(\frac{2j}{N})^2},~j=1,...,N
\end{eqnarray}
where $g$ defines an overall energy scale.
\begin{figure}
\includegraphics[width=.52\textwidth]{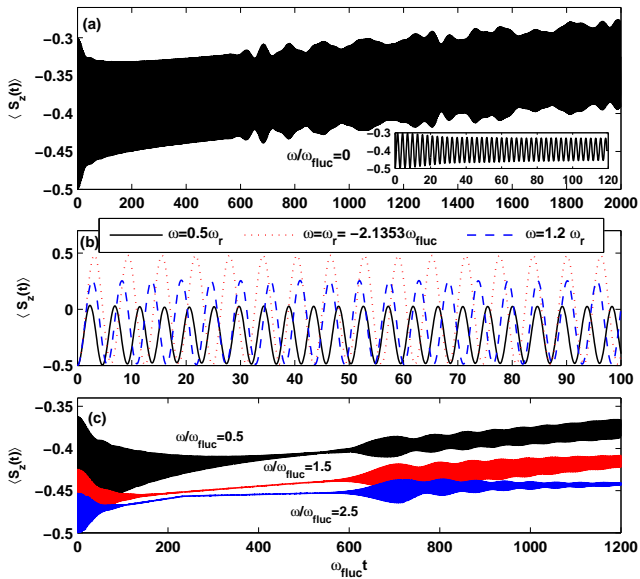}

  \caption{Polarization dynamics $\langle S_z(t)\rangle$ in the absence of intrabath interactions, $J/\omega_{\rm{fluc}}=0$. The bath is initially prepared in the fully polarized state Eq.~(\ref{fullp}) and  the hyperfine interaction is chosen to be isotropic with $\Lambda=1$. The number of bath spins is $N=30$. (a) In the absence of the magnetic field, $\omega/\omega_{\rm{fluc}}=0$. The inset shows the initial evolution up to $\omega_{\rm{fluc}}t=120$. (b) $\omega=0.5~\omega_r$, $\omega_r$, and $1.2~\omega_r$, where $\omega_r=-2.1353~\omega_{\rm{fluc}}$ is the resonant field defined in Ref.~\cite{pola3}. (c) For positive magnetic fields $\omega/\omega_{\rm{fluc}}=0.5,~1.5$, and $2.5$, the polarization dynamics shows collapse-revival behaviors.
  }
   \label{fig:1}
\end{figure}

\begin{figure}
  \includegraphics[width=.52\textwidth]{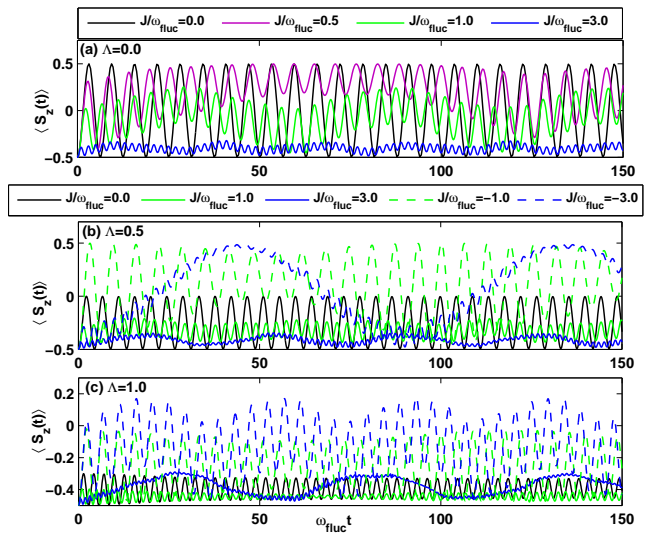}
  \caption{Polarization dynamics $\langle S_z(t)\rangle$ for different intrabath coupling $J$ and anisotropic factor $\Lambda$, with the bath initially prepared in the fully polarized state Eq.~(\ref{fullp}) and in the absence of the magnetic field, $\omega/\omega_{\rm{fluc}}=0$. The number of bath spins is $N=30$. (a) $\Lambda=0.0$ with $J/\omega_{\rm{fluc}}=0.0,~0.5,~1.0$, and $3.0$, (b) $\Lambda=0.5$ with $J/\omega_{\rm{fluc}}=0.0,\pm1.0$, and $\pm3.0$, (c) $\Lambda=1$ with $J/\omega_{\rm{fluc}}=0.0,\pm1.0$ and $\pm3.0$.}
   \label{fig:2}
\end{figure}
\begin{figure} 
(a) \includegraphics[width=.42\textwidth]{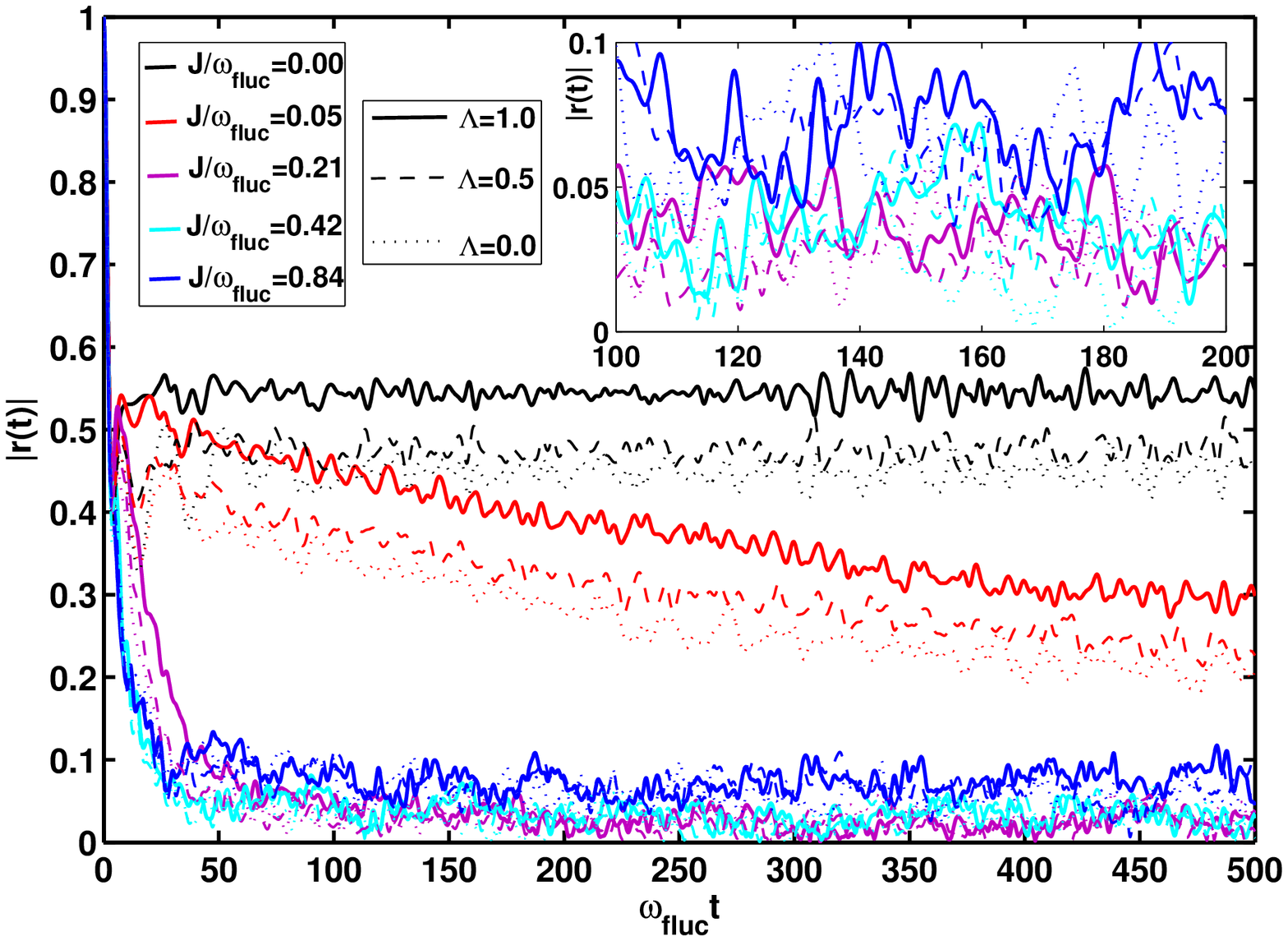}
(b)  \includegraphics[width=.43\textwidth]{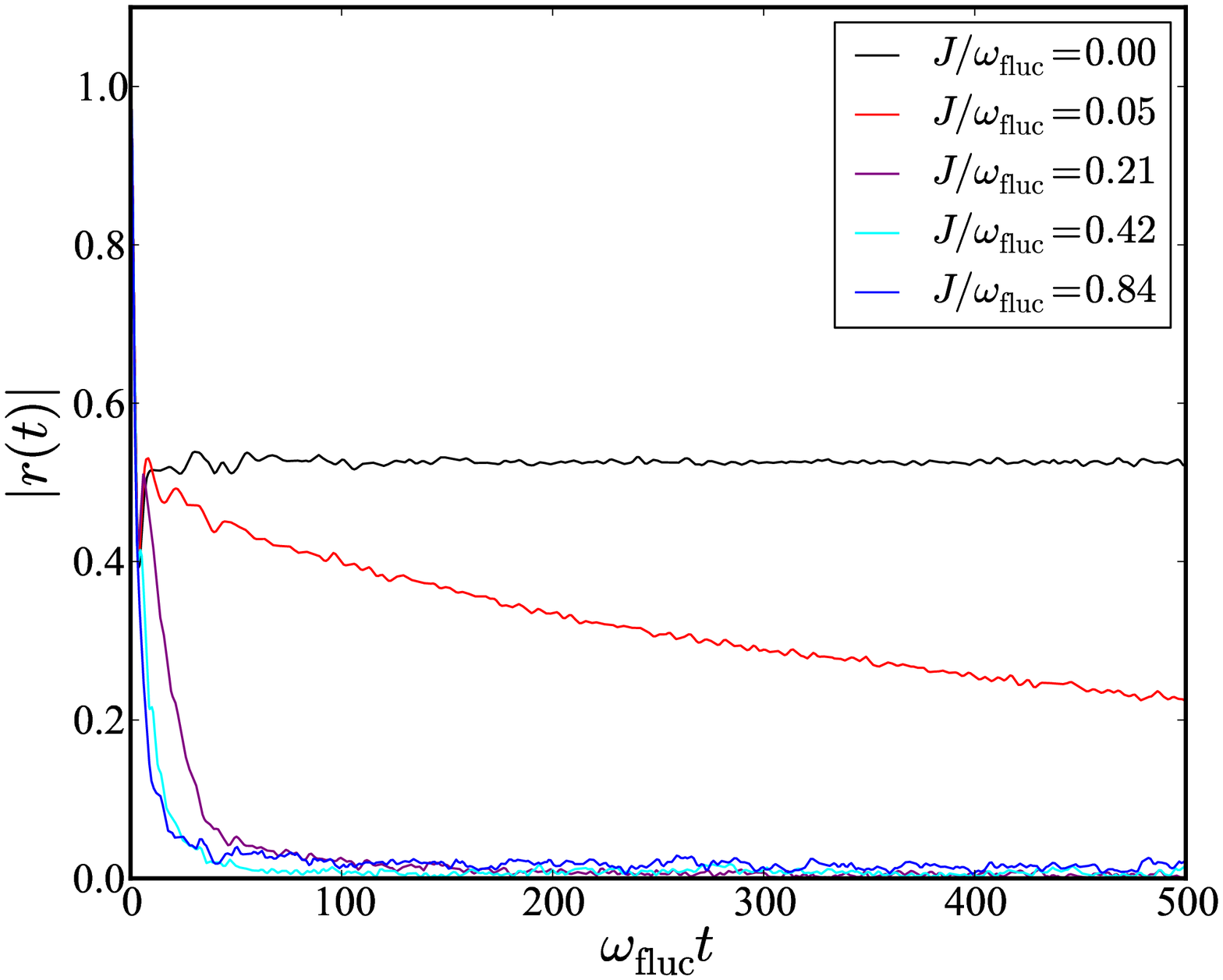}
(c) \includegraphics[width=.43\textwidth]{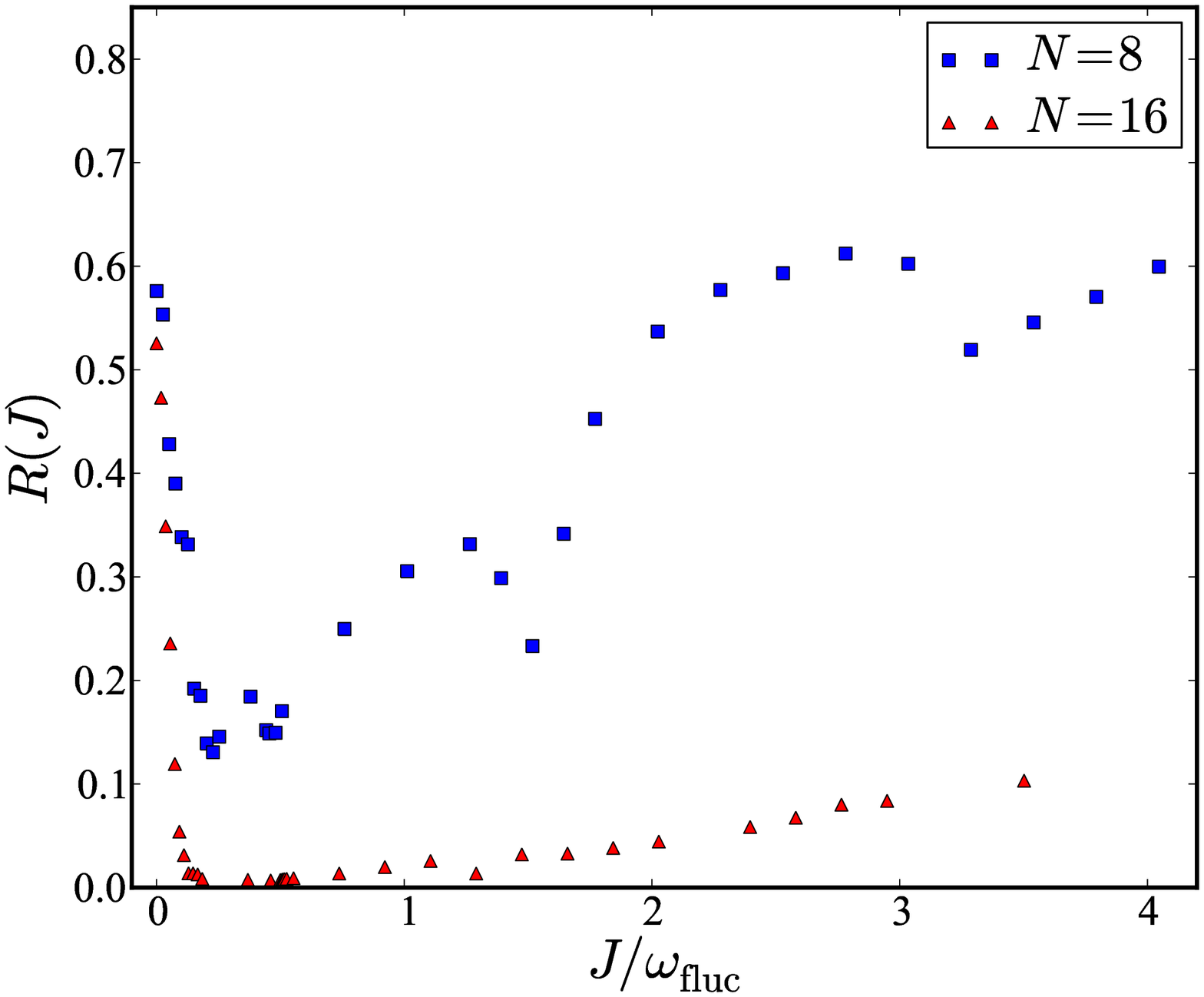}
\caption{(a) Real-time evolution of the decoherence factor $|r(t)|$ starting from the zero-polarization bath initial state Eq.~(\ref{zpb}) for $\omega=0$, $N=12$ bath spins using the EOM approach. The intrabath coupling strength $J/\omega_{\rm fluc}$ are chosen to cover both the weak coupling and strong coupling regime. The dashed and dotted curves present the corresponding decoherence dynamics for $\Lambda=0.5$ and $\Lambda=0.0$, respectively. The inset shows the magnification of the interval $\omega_{\rm fluc}T_{\max}\in[100,200]$.
 (b) $|r(t)|$ for the same initial conditions but obtained using the CET approach and $N=16$ bath spins, in the isotropic case $\Lambda=1$.
  (c) The long-time ($\omega_{\rm fluc}T_{\max}=700$) mean value $R(J)$ defined in Eq.\ \eqref{eq:40} versus $J/\omega_{\rm fluc}$ for $N=8$ and $16$, in the isotropic case $\Lambda=1$.}

 \label{fig:3}
\end{figure}
Let us first focus on the noninteracting and isotropic case with $J/\omega_{\rm{fluc}}=0$ and $\Lambda=1$. Fig.~\ref{fig:1}(a) shows the long-time dynamics of the polarization $\langle S_z(t)\rangle$ in the absence of the magnetic field, $\omega/\omega_{\rm{fluc}}=0.0$, obtained by solving Eq.~(\ref{C}) and Eq.~(\ref{A}) with initial conditions Eq.~(\ref{AC_ini}). The same result can be simply obtained by numerically integrating the equations of motion of real-space wavefunctions~\cite{pola1,pola2}. The inset of Fig.~\ref{fig:1}(a) shows the initial evolution, reproducing the result in Ref.~\cite{pola3} through numerically solving the Bethe ansatz equations for model (\ref{CSM}). For this case, we see that $\langle S_z(t)\rangle$ oscillates with a decreasing amplitude as time evolves, and the overall envelope finally decays in the long time limit. This is qualitatively consistent with Ref.~\cite{pola1,pola2}. Our results indicates that the analytic approximations derived from Bethe ansatz solutions~\cite{pola3} can correctly capture the polarization dynamics in the short or intermediate time regime, but may break down at sufficiently long time scales.
\par For negative magnetic fields, a resonant field around which exhibiting maximal oscillation amplitude and minimal frequency can be defined through $\omega_r =(-\sum_j g_j)\omega_{\rm{fluc}}$~\cite{pola3}. In Fig.~\ref{fig:1}(b) we plot $\langle S_z(t)\rangle$ for three different negative magnetic fields $\omega=0.5~\omega_r,~\omega_r$, and $1.2~\omega_r$. We see that consistent results with Ref.~\cite{pola3} are obtained. Interesting dynamics happens for positive magnetic fields. Fig.~\ref{fig:1}(c) shows the polarization dynamics for $\omega/\omega_{\rm{fluc}}=0.5$, $1.5$, and $2.5$. As the magnetic field increases, the Zeeman term $\omega S_z$ dominates and both the amplitudes of initial oscillations and the derivations of the overall profile from $\langle S_z(0)\rangle=-1/2$ decrease. Interestingly, collapse and revival behaviors~\cite{CR} arise for intermediate and strong magnetic fields that are larger than $\omega/\omega_{\rm{fluc}}\approx0.6$, during essentially the same period of time $\omega_{\rm{fluc}}t\in[0,700]$. No obvious collapse and revivals occur after $\omega_{\rm{fluc}}t\approx 700$ and the dynamics become less regular.
\par We also study effects of finite intrabath coupling $J$ and different anisotropic factors $\Lambda$ on the polarization dynamics. Fig.~\ref{fig:2}(a) presents $\langle S_z(t)\rangle$ for vanishing $\Lambda$, for which the model described by Eq.~(\ref{H}) is reduced to the model studied in Ref.~\cite{PRA}. Only results for positive $J$ are shown, since the dynamics is symmetric under the sign change of $J$. Fig.~\ref{fig:2}(b) and (c) show the polarization dynamics for $\Lambda=0.5$ and $\Lambda=1.0$, respectively. To compare results for different $\Lambda$'s, we use the same energy scale $\omega_{\rm{fluc}}$ for $\Lambda\neq1$. We observe that: i) for fixed $J$, the oscillation amplitude decreases, while the frequency increases as we increase the $\Lambda$. This means that the Ising component of the hyperfine interaction tends to increase the effective magnetic field on the central spin generated by the bath polarization, ii) for fixed $\Lambda$, amplitude increasing and frequency decreasing also occur for increasing $J$ to positive values. However, for large enough $J$ and finite $\Lambda$, fast oscillating of $\langle S_z(t)\rangle$ tends to establish an envelope slowly oscillating around a mean value, iii) In the case of $\Lambda\neq0$, changing the sign of $J$ to negative values will increase the overall amplitude of the oscillation profile to regions of $\langle S_z(t)\rangle>0$.


\subsection{Zero-polarized antiferromagnetic bath state}

We now turn to the study of the central spin decoherence dynamics with the bath initially prepared in a zero-polarization antiferromagnetic state~\cite{mc1},
 \begin{eqnarray}\label{zpb}
 |\phi_b\rangle&=&|\bar{1}1\bar{1}...\bar{1}1\bar{1}\rangle=\sum_{\vec{k}_{\frac{N}{2}}}S^*_{\vec{k}_{\frac{N}{2}};(2,4,...,N)}  |\vec{k}_{\sigma,\frac{N}{2}}\rangle.
 \end{eqnarray}
yielding the initial values for the $\chi's$
 \begin{eqnarray}
\chi_{\vec{k}_{\frac{N}{2}}}=S^*_{\vec{k}_{\frac{N}{2}};(2,4,...,N)}.
 \end{eqnarray}
For this type of initial state, we will use an exponential distribution of coupling constants \cite{mc1}
 \begin{eqnarray}
 g_j=\frac{g}{N}e^{-\frac{j-1}{N}},
 \end{eqnarray}
where
$g$ defines the overall energy scale. In this case, $\omega_{\rm{fluc}}$
is given by
\begin{eqnarray}
\omega_{\rm{fluc}}=\frac{2g}{N}e^{\frac{1}{N}-1}\sqrt{\frac{ e^2-1 }{ e^{\frac{2}{N}}-1}}.
 \end{eqnarray}
The nonperturbative crossover regime and strong magnetic field regime are defined by $\omega_{\rm{fluc}}\lesssim\omega\lesssim 2g$ and $\omega>2g$, respectively~\cite{mc1}.

Here the main quantity of interest is the decoherence factor $|r(t)|$ of a system initially prepared in a superposition initial state $a_1|1\rangle+a_{\bar{1}}|\bar{1}\rangle$ for the central spin.
Conservation of the total magnetization restricts the bath polarization to the
three sectors  with $m=\frac{N}{2}-1,\frac{N}{2}$, and $\frac{N}{2}+1$, independent of $J$.
The large number of $f$-functions ($\sim (C^{N/2}_N)^2$) to be calculated prevents us from dealing with baths even with intermediate number of spins.
Therefore, we limit ourselves to $N\leq14$ in the following numerical calculations using the EOM approach.

\par We have performed benchmark
calculations in small systems by simulating the decoherence dynamics of the central spin using the two different methods. Fig. \ref{fig:3}(a) and (b) show the time evolution of the decoherence factor $|r(t)|$ for several different values of the intrabath coupling $J$ at zero magnetic field, $\omega=0$, with the number of bath spins $N=12$ and $16$, respectively. Since the EOM and CET results are identical for $N=12$,
Fig. \ref{fig:3}(a) presents the data for both methods while we have employed only the CET approach for
the results in Fig. \ref{fig:3}(b) for $N=16$. The intrabath coupling strength $J$ are chosen to cover both the weak and strong coupling regime, by noting that the largest hyperfine coupling (in the isotropic case) $2g/N\approx0.42~\omega_{\rm fluc}$ and $0.37~\omega_{\rm fluc}$ for $N=12$ and $16$, respectively.

\par In Fig. \ref{fig:3}(a), we also present results for anisotropic hyperfine couplings with $\Lambda=0.5$ and $\Lambda=0.0$. We see that, for relatively small $J$, increasing of the Ising component of the hyperfine coupling strength tends to enhance the coherence. We also calculated $|r(t)|$ for negative $J$'s (not shown), and find no essential deviation of the overall profiles from that of positive $J$'s. This might be due to the fact that an overall sign change of the bath Hamiltonian does not change the energy gaps of the bath. In the following discussion, we will focus on the isotropic case, $\Lambda=1$.

The effect of a non-vanishing  $J\neq0$ is significant.
For all values of $J$ considered, $|r(t)|$ drops off
from the initial value of $|r(0)|=1$ on the characteristic short time scale $1/\omega_{\rm fluc}$
set by the fluctuation of the Overhauser field and tends to oscillate around a fixed average value. Consistency with Ref.~\cite{mc1} is found for $J=0$,
where $|r(t)|$ oscillates about a mean value of $\approx 0.55$ at immediate time scales
for $N=12$. With increasing $N$, this mean value (defined below in Eq.~(\ref{eq:40})) gradually decreases as can be seen by comparing the $J=0$ cases in Fig. \ref{fig:3}(a) and \ref{fig:3}(b), as well as from Fig. \ref{fig:3}(c). We expect that it will reach  $1/2$ as $N\to \infty$
if the long-time average of the coefficients $|A|$ and $|B|$ remain finite
and the initial spectral weight is distributed equally amongst $A,B,C$, and $D$.

Switching on the bath interaction $J$ introduces additional spin-flip processes in the
spin bath. After such a spin-flip process occurs between two bath spins, however,
the Ising contribution of the hyperfine interaction will change sign leading to a destructive interference of
the terms in  Eq.\ \eqref{eq:r-vs-t}.  Consequently, the long-time limit of $|r(t)|$ decreases
with increasing $J$ starting from $J=0$ as can be observed  in Fig. \ref{fig:3}(a) and (b).

The other extreme regime of our model is  the limit of very large $J$.
In this case, the initial state of the bath $|\phi_b\rangle$ is expanded in
the eigenstates of the $XX$ ring, $|\vec{k}_{\sigma,m}\rangle$ with fixed $m=N/2$,
and the hyperfine coupling to the central spin is considered as a weak
perturbation.  For a finite $N$, the spectrum is discrete and
the differences between different eigenenergies are increasing with $J$.
Above some value of $J$, even the smallest excitation energy exceeds the largest energy scales
of hyperfine coupling, $g_1$, and spin-flip processes
corresponding to a  creation of a particle or hole excitation
will be suppressed.  Consequently, the long-time limit of $|r(t)|$ increases.
For a system with no degeneracies, full coherence, i.\ e.\ $|r(t)|=1$,
would be restored, and quantum coherence is maintained at all times.
However, spin-flip processes can couple energetically degenerate eigenstates
with $m$ and $m\pm1$. One can find different combinations of $\vec{k}_m$ and $\vec{k}_{m\pm 1}$ such
that their corresponding occupation energies are the same. However, since the dispersion for $N_f$ even and odd are different, this coincident energy can only be $\mathcal{E}_{\vec{k}_m}= \mathcal{E}_{\vec{k}_{m\pm1}}=0$. Independent of the strength of the perturbation, $g_j/J$, the degeneracy is lifted upon switching
on the hyperfine coupling,  and
the states are maximally entangled. The arbitrarily weak spin-flip processes
generated a mixture of local spin up and down states, and, therefore,
the mean values of $|r(t)|$ reaches a plateau less than unity for $J\to \infty$.
This plateau value is dependent on $N$.
One may naively think that this plateau value will decrease with increasing $N$ due to the increasing degree of
degeneracies of coupled subsectors. However, our numerical simulation indicates that this is not the case, as we show below.

\begin{figure}[tb]
  \includegraphics[width=.40\textwidth]{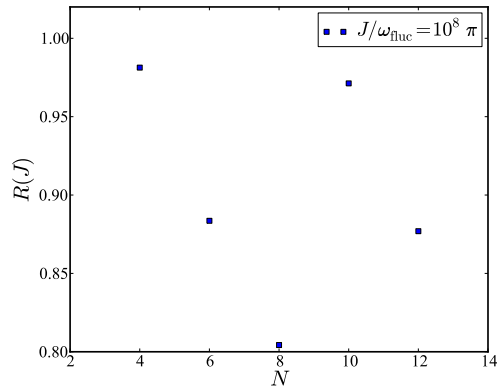}
 \caption{Plateau value of $R(J)$ as function of $N$  for $J/\omega_{\rm fluc}\gg 1$
 calculated at  a fixed  $J/\omega_{\rm fluc}=10^8\pi$.
}
\label{fig:P-J-plateau}
\end{figure}

To this end, the decoherence is increasing with increasing $J/\omega_{\rm fluc}$
starting from $J=0$, and the coherence is partially restored in a finite size system,
for $J\to \infty$ due to the entanglement in energetically degenerate subsectors.
Consequently, there must be a value $J/\omega_{\rm fluc}$ where the long-time mean value
of the decoherence factor $|r(t)|$ approaches a minimum. In order to quantify this point
we have  defined the time averaged $R(t)$
\begin{eqnarray}
\label{eq:40}
R(t,J) &=& \frac{1}{t-t_{\rm min}} \int_{t_{\rm min}}^t d\tau |r(\tau,J)|
\end{eqnarray}
for $t> t_{\rm min}$ in order to eliminate the finite-size oscillations of
$|r(t)|$ in the long-time limit for different $J$. To separate the initial short-time decay
of $|r(t)|$ from its long-time mean values, we start the integration
at $\omega_{\rm fluc}t_{\rm min}= 100$.

We define the $J$ dependent long-time limit as $R(J)=R(T_{\rm max},J)$.
The results for $R(J)$ at the largest time scale of our simulation, $\omega_{\rm fluc}T_{\rm max}=700$,
are depicted in Fig. \ref{fig:3}(c) as function of $J/ \omega_{\rm fluc}$ for $N=8$ and $16$.
For $J=0$ we note the predicted decrease of $R(J=0)$  with increasing $N$ towards $1/2$.
Surprisingly,  for $N=8$, $R(J)$ vs $J$ shows not only a single, but several very sharp local minima: the first two
at $J/\omega_{\rm fluc}=0.28$ and $0.47$. This is a clear indicator of finite-size effects: at some values of $J$ the decoherence is enhanced,
which might be related to an increase of the degrees of degeneracy.

For $N=16$, $R(J)$ does not exhibit such pronounced local minima, almost vanishes at $J/\omega_{\rm fluc}\approx 0.23$ and slowly increases beyond that point. The number of eigenstates of the system are exponentially larger than for $N=8$, and, therefore,
the mean level spacing of the excitation spectrum is considerably smaller. Nevertheless,
$R(J)$ increases again for larger $J$ after passing the minimum.

We always find that $R(J)$ approaches a finite mean value for large $J$.
In Fig.\ \ref{fig:P-J-plateau}, $R(J)$ is plotted as function of $N=4,6,8,10$, and $12$ at $J/\omega_{\rm fluc}=10^8\pi\gg 1$.
We find an oscillatory behavior with $N/2$ changing its parity. In order to qualitatively understand this behavior, we note that there are totally $\mathcal{N}^{(m)}_N=C^{m}_N$ eigenstates in the $m$-sector of the bare $XX$ ring. Among these, the number of states with zero energy is denoted by $\mathcal{M}^{(m)}_N$. Then the spin-flip induced decoherence can be roughly measured by the following \emph{degeneracy factor}
\begin{eqnarray}
\label{df}
\Theta^{(m)}_N\equiv\frac{\mathcal{M}^{(m)}_N}{\mathcal{N}^{(m)}_N}\frac{\mathcal{M}^{(m-1)}_N}{\mathcal{N}^{(m-1)}_N}=\frac{\mathcal{M}^{(m)}_N}{\mathcal{N}^{(m)}_N}\frac{\mathcal{M}^{(m+1)}_N}{\mathcal{N}^{(m+1)}_N}.
\end{eqnarray}
The larger $\Theta^{(m)}_N$, the large the decoeherence. Although $\mathcal{N}^{(N/2)}_N$ is a regular function of $N$, the number of degenerate states $\mathcal{M}^{(N/2)}_N$ varies with $N$ less regularly, especially for odd $N/2$, due to different spectrum structures for $N/2=$ even or odd. For example, $\mathcal{M}^{(3)}_6$, $\mathcal{M}^{(5)}_{10}$, and $\mathcal{M}^{(7)}_{14}$ all equal to $2$, but $\mathcal{M}^{(9)}_{18}=692$. Calculating $\Theta^{(N/2)}_N$ for $N=6, 8, 10$, and $12$ we get $\Theta^{(3)}_6=3.33\times 10^{-2}$, $\Theta^{(4)}_8=4.59\times 10^{-2}$, $\Theta^{(5)}_{10}=9.83\times 10^{-4}$, and $\Theta^{(6)}_{12}=7.62\times 10^{-3}$, which basically reveal the oscillating behavior of $R(J)$ with increasing $N$. The details, however, will depend on the matrix elements. Since we recover $R(J)=1$ for $g_j=0$, the decoherence is discontinuous for $g_j=0$ to finite $g_j>0$ which is related to lifting of degeneracies in the interacting spin bath and maximizing the entanglement in this degenerate subsection by switching on $g_j$ independent of its values.
\par Although the degeneracy argument presented here rests on a translational bath for which the degeneracy factor can be easily calculated, it still applies for cases with nonuniform intrabath coupling $\{J_j\}$ ($J_j$ is the intrabath coupling strength between spin-$j$ and spin-$(j+1)$ in the bath). In general, the degeneracies within a certain magnetization sector will be lifted by introducing inhomogeneity of $\{J_j\}$. However, the degeneracy argument still works since it only relies on the perturbative consideration at large $J$. As an illustration, we give in the Appendix an example study of a simple system having only $N=4$ bath spins, but with inhomogeneous intrabath coupling $\{J_1,J_2,J_3,J_4\}$.
\par The numerical calculations in this work are limited to relatively small baths with $N\le16$ bath spins. However, for the dynamics of the conventional central spin model, it is usually believed that $N=16$ is large enough to eliminate finite-size effects and qualitatively similar behaviors with larger $N$ can be seen. Thus, we belived that the above observations, namely the existence of the optimal $J=J_{\rm opt}$ that enhances the decoherence the most, and the increase of $|r(t)|$ when $J$ is passing this value should remain for large baths with $N>16$. As a result of the interplay between the intrabath coupling and the spin-bath coupling, the minimum of $R(J)$ at $J_{\rm opt}$ occurs when $J$ is comparable to $\{g_j\}$. We note that the strongest dimensionless spin-bath coupling $2g/(N\omega_{\rm fluc})$ decreases as $N$ increases, which will move $J_{\rm opt}/\omega_{\rm fluc}$ to the left.

\section{Conclusions and discussions}
In this work we studied the real-time quantum dynamics of a generalized central spin model, which extends the usual central spin model or Gaudin model by arranging the bath spins on a circle and introducing nearest-neighbor homogeneous $XX$-type intrabath interactions. This model can also be viewed as an extension of the model studied in our previous work~\cite{PRA} by including the Ising part of the system-bath coupling. Taking advantage of the conservation of total magnetization and translational invariance of the resulting spin bath, we could work in the bath's momentum space and write down the most general forms of the time-dependent wavefunctions. The exact equations of motion for the corresponding probability amplitudes in the momentum space are derived and integrated numerically to obtain the time evolution of the polarization $\langle S_z(t)\rangle$ and decoherence factor $|r(t)|$ of the central spin. It can be derived from the EOMs that the initial decay of $|r(t)|$ will start from at least the quadratic term of time $t$. For the case of uniform system-bath coupling, the EOM framework gives a transparent physical picture based on the momentum-conserving scattering processes into adjacent $L_z$-subsectors of the bath.
\par Using the above results, we first examined the central spin polarization dynamics with the bath initially prepared in a fully polarized state. This is followed by the calculation of the decoherence dynamics of the central spin starting from an antiferromagnetic bath initial condition, where the high dimensions of the Hilbert spaces involved prevent us from dealing with large baths with $N\geq16$ spins by using the EOM method. However, thanks to the recently developed Chebyshev expansion technique which proves its high efficiency in simulating the dynamics of the conventional central spin model~\cite{anders}, we could apply the CET approach to the current model to treat relatively large baths as well as ultrastrong intrabath interactions. Consistence with prior works was found in the absence of the intrabath coupling. We revealed novel dynamical behaviors of the decoherence for finite intrabath coupling strengths. In general, the decoherence factor $|r(t)|$ drops off abruptly at the initial stage and approaches a roughly steady value at long times. To illustrate this, we introduced a time-dependent quantity $R(t,J)$ to describe the time averaged value of $|r(t)|$ at a given intrabath coupling strength $J$. Intriguingly enough, we found that the long-time mean value $R(J)=R(T_{\max},J)$ shows a non-monotonic dependence on $J$. For large enough baths with vanishing finite-size effect, there always exists some value of $J$ that enhances the decoherence the most. However, as $J$ increases further by passing this value, the coherence is restored and tends to be some fixed value that depends on the number of bath spins $N$. This dependence shows an oscillatory behavior as $N/2$ changes its parity and can be understood by degeneracy analysis of states in different magnetization sectors based on the EOMs and perturbation theory.

\par Finally, we would like to discuss possible experimental realizations of our model in nanoscale systems. Consider a modified `quantum corral'~\cite{corral1} construction by coating a metallic substrate with a thin insulating layer, on the top of which one places iron adatoms in a ring structure. The presence of the insulating layer will cut off the Kondo effect~\cite{corral2} induced by the conduction electrons in the metallic host, so that a direct Heisenberg type exchange interaction between neighboring adatoms might be left. By further placing an additional single iron adatom in the inner of the quantum corral, a model with Heisenberg type intrabath as well as system-bath couplings can be realized. The intrabath coupling strength $J/\omega_{\rm{fluc}}$ can reach the strong coupling regime reasonably due to the generally larger radius of the corral compared to the distance between adjacent bath atoms. Although our model only includes an $XX$-type intra-ring interaction, we believe it is a first step towards understanding this realistic system. Furthermore, the CET technique can, in principle, also deal with the dynamics of such a system with Heisenberg type intrabath coupling~\cite{prog}.

\par Our analysis implies that intrabath coupling could have significant effects on the central spin decoherence. The interacting central spin model and theoretical methods introduced by us may stimulate further studies on the intrabath interaction effects in more realistic solid-state central spin systems.

\noindent{\bf Acknowledgements:}
NW thanks N.~A.~Sinitsyn for useful discussions and critical reading of the manuscript. We acknowledge support from NSF Grant No. CHE-1058644 (NW) and ARO-MURI Grant No. W911NF-13-1-0237 (HR). Some of us (HJ, NF, FBA)  acknowledge the financial support by the Deutsche Forschungsgemeinschaft
and the Russian Foundation of Basic Research in the frame of the ICRC TRR 160.

\appendix
\section*{Appendix: A 4-spin inhomogeneous bath}
In this appendix, we will calculate the decoherence dynamics $|r(t)|$ for a spin bath made up of four bath spins with inhomogeneous intrabath couplings. The system is described by
\begin{eqnarray}
H_4=\sum^4_{j=1}J_j(S^x_jS^x_{j+1}+S^y_jS^y_{j+1})+2\sum^4_{j=1} g_j \mathbf{S}\cdot \mathbf{S}_j.
\end{eqnarray}
As in Section III.B, the initial state is chosen as $|\psi(0)\rangle=\frac{1}{\sqrt{2}}(|1\rangle+|\bar{1}\rangle)|\bar{1}1\bar{1}1\rangle$. Diagonalizing the spin bath within the $m=2$ (spanned by $\{ |11\bar{1}\bar{1}\rangle, |1\bar{1}1\bar{1}\rangle, |1\bar{1}\bar{1}1\rangle, | \bar{1}11\bar{1}\rangle, | \bar{1}1\bar{1}1\rangle, | \bar{1}\bar{1}11\rangle\}$) and $m=3$ (spanned by $\{|111\bar{1}\rangle,|11\bar{1}1\rangle,|1\bar{1}11\rangle,|\bar{1}111\rangle\}$) subsectors gives the following eigenvalues,
\begin{eqnarray}
E^{(m=2)}_{\sigma,\sigma'}&=& \frac{\sigma}{2}\sqrt{\sum_jJ^2_j+2\sigma'(J_1J_3+J_2J_4)},\nonumber\\
E^{(m=2)}_{5}&=&E^{(m=2)}_{6}=0,
\end{eqnarray}
and
\begin{eqnarray}
E^{(m=3)}_{\sigma,\sigma'}&=& \frac{\sigma}{2\sqrt{2}} \sqrt{\sum_jJ^2_j+\sigma'\sqrt{(\sum_jJ^2_j)^2-4(J_1J_3-J_2J_4)^2}},\nonumber\\
\end{eqnarray}
where $\sigma,\sigma'=\pm 1$. We see that there always exist two zero-energy eigenstates in the $m=2$ sector. For the homogeneous coupling $J_j=J$, we have $E^{(m=2)}_{\pm,1}=\pm\sqrt{2}|J|,~E^{(m=3)}_{\pm,1}=\pm |J|$, and $E^{(m=2)}_{\pm,-1}=E^{(m=3)}_{\pm,-1}=0$, so that the four eigenstates with zero energy $E^{(m=2)}_{\pm,-1},E^{(m=2)}_{5,6}$ in the $m=2$ sector, and the two eigenstates with zero energy $E^{(m=3)}_{\pm,-1}$ in the $m=3$ sector mainly participate the mixing of the two subsectors. Introducing inhomogeneity in $\{J_j\}$ will generally lift these degeneracies. However, if the energy splitting is of the order of $\{g_j\}$, fast decay of $|r(t)|$ is still possible. 
\begin{figure} 
  \includegraphics[width=.50\textwidth]{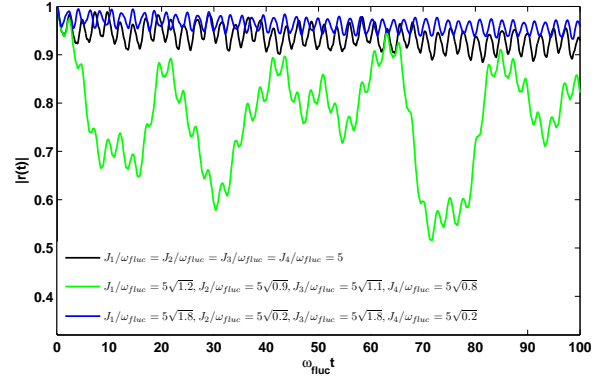}
 \caption{Decoherence dynamics $|r(t)|$ of a central spin coupled to an interacting spin bath with $N=4$ bath spins. Black: $J_1=J_2=J_3=J_4=J$; green: $J_1=J\sqrt{1.2},J_2=J\sqrt{0.9}, J_3=J\sqrt{1.1}, J_4=J\sqrt{0.8}$; blue: $J_1=J\sqrt{1.8},J_2=J\sqrt{0.2}, J_3/\omega_{\rm fluc}=J\sqrt{1.8}, J_4=J\sqrt{0.2}$. The three sets of parameters are chosen such that $\sum_jJ_j^2$ is unchanged. Here $J=5\omega_{\rm fluc}$ is the homogeneous intrabath coupling.}
\label{fig:app}
\end{figure}
\par In Fig.~(\ref{fig:app}), we plot $|r(t)|$ for three sets of intrabath coupling configurations $\{J_j\}$, which all lie in the strong intrabath coupling regime in the sense that the smallest $J_j/\omega_{\rm fluc}\gg 2g_1/\omega_{\rm fluc}=0.6746$ for $J/\omega_{\rm fluc}=5$. We keep $\sum_jJ_j^2/4=J^2$ as a constant for each set of the parameters. For $J_1=J\sqrt{1.2},J_2=J\sqrt{0.9}, J_3=J\sqrt{1.1}, J_4=J\sqrt{0.8}$ (green curve), we see that $|r(t)|$ drops further compared with the homogeneous case. This can be understood by looking at the eigenvalues of the spin bath: $E^{(m=2)}_{\pm,1}=\pm7.0688~\omega_{\rm fluc}$, $E^{(m=2)}_{\pm,-1}=\pm0.1789~\omega_{\rm fluc}$, and $E^{(m=3)}_{\pm,-1}=\pm0.3765~\omega_{\rm fluc}$, $E^{(m=3)}_{\pm,1}=\pm4.9858~\omega_{\rm fluc}$. All the energy splittings are of the order of $\{g_j\}$. In contrast, suppression of the decay of $|r(t)|$ is observed for $J_1=J\sqrt{1.8},J_2=J\sqrt{0.2}, J_3=J\sqrt{1.8}, J_4=J\sqrt{0.2}$ (blue curve). In this case, the spectrum in the $m=2$ sector does not change as $J_1J_3+J_2J_4=2J^2$. However, $E^{(m=3)}_{\pm,-1}$ raise up to $\pm2.2361~\omega_{\rm fluc}\gg g_j$, which consequently suppress the mixing of the two subsectors.

\end{document}